\documentclass[usenatbib]{mn2e}
\usepackage{amsmath,amssymb,graphicx,aas_macros}

\begin{document}
\title{The Clustering and Host Halos of Galaxy Mergers at High Redshift}
\author[Wetzel, Cohn \& White]{Andrew R. Wetzel${}^{1}$, J.D. Cohn${}^{2}$, 
Martin White${}^{1,3}$\\
$^{1}$Department of Astronomy, University of California, Berkeley, CA 94720, 
USA\\
$^{2}$Space Sciences Laboratory, University of California, Berkeley, CA 94720, 
USA\\
$^{3}$Department of Physics, University of California, Berkeley, CA 94720, USA
} 

\date{October 2008}

\pagerange{\pageref{firstpage}--\pageref{lastpage}} \pubyear{2008}

\maketitle

\label{firstpage}

\begin{abstract} 
High-resolution simulations of cosmological structure formation indicate that 
dark matter substructure in dense environments, such as groups and clusters, 
may survive for a long time.
These dark matter subhalos are the likely hosts of galaxies.
We examine the small-scale spatial clustering of subhalo major mergers at high 
redshift using high-resolution N-body simulations of cosmological volumes.
Recently merged, massive subhalos exhibit enhanced clustering on scales 
$\sim 100-300 h^{-1}\,$kpc, relative to all subhalos of the same infall mass, 
for a short time after a major merger ($<500\,$Myr).
The small-scale clustering enhancement is smaller for lower mass subhalos, 
which also show a deficit on scales just beyond the excess.
Halos hosting recent subhalo mergers tend to have more subhalos; for massive 
subhalos the excess is stronger and it tends to increase for the most massive 
host halos.
The subhalo merger fraction is independent of halo mass for the scales we probe.
In terms of satellite and central subhalos, the merger increase in small-scale 
clustering for massive subhalos arises from recently merged massive central 
subhalos having an enhanced satellite population.
Our mergers are defined via their parent infall mass ratios.
Subhalos experiencing major mass gains also exhibit a small-scale clustering 
enhancement, but these correspond to two-body interactions leading to two final 
subhalos, rather than subhalo coalescence.
\end{abstract}


\section{Introduction}

A wealth of high redshift galaxy data is now accumulating, and many members in 
the resultant galaxy zoo are thought arise from galaxy mergers: quasars 
\citep{Car90}, Lyman Break Galaxies \citep[LBG; see][for review]{Gia02}, 
submillimeter galaxies \citep[SMG; see][for review]{Bla02}, ultra-luminous 
infrared galaxies \citep[ULIRG; see][for review]{SanMir96} and starburst and 
starburst remnant galaxies \citep[e.g.,][]{BarHer91,Nog91}.
Observational samples of these objects at $z\gtrsim 1$ are growing large enough 
to produce statistical measurements of clustering which can be compared to 
candidates in numerical simulations \citep[e.g.,][]{Gia98,Bla04,Cro05,Ouc05,
CooOuc06,Hen06,Kas06,Lee06,ScoDunSer06,Coi07,Gaw07,She07,Fra08,Mye08,
YamYagGot08,Yan08,Yos08}.

In high resolution dark matter simulations, over-dense, self-bound, dark matter 
substructures in dense environments can survive for a long time \citep{Tor97,
TorDiaSye98,Ghi98,KlyGotKra99,Moo99}.
These ``subhalos'' are thought to be the hosts of galaxies, and indeed the
identification of galaxies with subhalos reproduces many galaxy properties
\citep[e.g.,][]{Spr01,Spr05,ZenBerBul05,Bow06,ConWecKra06,ValOst06,WanKauDeL06}.
The complex dynamics of subhalos may thus be a good proxy for those of galaxies
themselves, suggesting that galaxy mergers can be identified with subhalo  
mergers within simulations.
We will use the terms galaxy and subhalo interchangeably hereon.
Measurements of subhalo mergers can provide a quantitative reference for the 
identification of merger related objects in observations and can aid in the 
correct interpretation of their clustering measurements.

One question of particular interest is whether small-scale clustering can probe 
merger activity.
There has been much recent discussion of a ``small-scale clustering 
enhancement'' of galaxy subpopulations in simulations and observations, but 
several differing definitions of ``enhancement'' exist, mostly stemming from 
different choices of reference.
For instance, small-scale enhancement has been used to describe clustering 
stronger than a power law extrapolated from larger scales, or clustering 
stronger than that of dark matter at small scales.
However both of these behaviors are seen in non-merging samples.
Luminous Red Galaxies in SDSS are not thought to be associated with (recent)
mergers, yet their correlation function is much steeper than the dark matter on 
scales of several hundred kpc \citep{Mas06}.
One expects objects which populate halos more massive than $M_\star$ (the
characteristic non-linear mass) will have an upturn in their correlation
function on scales below the virial radius of the $M_\star$ halos since at this
scale the clustering is dominated by pairs of objects within the same halo and 
the mass function is very steep \citep[e.g.,][]{Sel00,PeaSmi00}.
Quasars (thought to be associated with mergers) do exhibit a sharp upturn in
clustering on very small scales ($25-50\,h^{-1}$kpc) at $z \approx 1-3$
\citep{Hen06,Mye08}, but a comparison with galaxy clustering measurements at
these scales and redshifts is lacking.\footnote{
It is not clear how to interpret a galaxy clustering measurement on scales 
smaller than the galactic radius.}
On slightly larger scales, low $z$ quasar clustering observations show no excess
above a power law \citep[e.g.][and discussion therein]{Pad08}.

On the other hand, merger related clustering effects are not unexpected.
Recently, various authors have shown that the large-scale clustering of dark
matter halos depends on their formation histories 
\citep[known as ``assembly bias''; e.g.,][]{SheTor04,GaoSprWhi05,Wec06}, and in 
particular, on recent halo merger activity \citep[``merger bias'';][]{ScaTha03,
Wet07,WetSchHol08}.\footnote{
See also \protect\citet{FurKam06} for analytic estimates, \protect\citet{Per03}
for a simulation limit on the effect, \protect\citet{CroGaoWhi06,Tin08} for 
halo assembly bias applied to galaxy clustering.}
Any such history or merger dependent clustering breaks the usually assumed 
direct link between large-scale clustering amplitude and halo mass.
Analytical modeling of the small-scale clustering of quasars has been compared 
to observations, assuming that quasars are mergers and that mergers occur in 
denser environments \citep{HopHerCox08}.

In this paper, we focus on the small-scale clustering of subhalo mergers and 
its relation to their host halos.
We use dark matter simulations to compare the small-scale clustering of 
recently merged subhalos at high redshifts to the clustering of the general 
subhalo population of the same infall mass.
To interpret our results, we relate the merged subhalos to their host dark 
matter halos using the formalism of the halo model \citep{Sel00,PeaSmi00,
CooShe02}, examining the dependence of subhalo mergers both on their host halo 
masses and their halo radial distribution profiles.

In \S\ref{sec:numerical} we briefly describe the simulations and summarize the 
main properties of subhalo tracking and merger definitions as detailed
in \citet{WetCohWhi09a}, hereafter called Paper I.
In \S\ref{sec:xi} we compare the small-scale clustering of the recently merged 
and full population of subhalos.
We also describe results for the alternate mass gain merger definition.
In \S\ref{sec:hod} we use the framework of the halo model to identify 
contributions to merger clustering and to understand the relation of subhalo 
mergers to their host dark matter halos.
We note properties of merger pairs in \S\ref{sec:pairs}, and we summarize and 
discuss our results in \S\ref{sec:discussion}.

\section{Numerical Techniques and Merger Definitions} 
\label{sec:numerical}

\subsection{Simulations and Subhalo Tracking}

Our simulation and subhalo finding and tracking details are discussed 
extensively in Paper I; we summarize the main features here.
We use two dark matter only N-body TreePM \citep{TreePM} simulations of $800^3$ 
and $1024^3$ particles in periodic cubes with side lengths $100\,h^{-1}$Mpc 
and $250\,h^{-1}$Mpc, respectively.
For our $\Lambda$CDM cosmology ($\Omega_m=0.25$, $\Omega_\Lambda=0.75$, 
$h=0.72$, $n=0.97$ and $\sigma_8=0.8$), in agreement with a wide array of 
observations \citep{COBE,Teg06,ACBAR,WMAP5}, this results in particle masses of 
$1.4\times 10^8h^{-1}M_\odot$ ($1.1 \times 10^9 h^{-1}M_\odot$) and a 
Plummer equivalent smoothing of $4\,h^{-1}$kpc ($9\,h^{-1}$kpc) for the smaller 
(larger) simulation.
Outputs were spaced every $50\,$Myr ($\sim 100\,$Myr) for the smaller (larger) 
simulation, from $z\sim 5$ to $2.5$.
Additional outputs from the smaller simulation were retained at lower redshift,
spaced every $\sim 200\,$Myr down to $z=0.6$, below which we no longer fairly 
sample a cosmological volume.

We find subhalos (and sometimes subhalos in subhalos) by first generating a 
catalog of halos using the Friends-of-Friends (FoF) algorithm \citep{DEFW} with 
a linking length of $b=0.168$ times the mean inter-particle 
spacing.\footnote{
The longer linking length of $b=0.2$ is often used, but it is more susceptible 
to joining together distinct, unbound structures and assigning a halo that 
transiently passes by another as a subhalo.
Halos at our mass and redshift regime have roughly a $\sim 15\%$ lower FoF mass 
than for $b=0.2$.}
We keep all groups that have more than $32$ particles, and halo masses quoted 
below are these FoF masses.
Within these ``(host) halos'' we then identify ``subhalos'' as gravitationally 
self-bound aggregations of at least $20$ particles bounded by a density saddle 
point, using a new implementation of the {\sl Subfind\/} algorithm 
\citep{Spr01}.
The central subhalo is defined as the most massive subhalo within its host halo,
and it includes all halo matter not assigned to satellite subhalos.\footnote{
Since {\sl Subfind\/} defines the central subhalo as the most massive subhalo, 
it is not necessarily the case that the central subhalo contains the most 
dense or deepest potential particles.}
Thus, the central subhalo mass represents halo mass not bound to any satellite
subhalos and so it should track the gas mass available to accrete onto the 
central galaxy.
Subhalo positions are those of their densest particle, and the halo position is
that of its central suhalo.

Each subhalo is given a unique child at a later time, based on its $20$ most 
bound particles.
We track subhalo histories across four consecutive outputs at a time since 
subhalos can briefly disappear during close passage with another subhalo, 
i.e. ``fly-by's''.
Additionally, at these redshifts, the distinction between a central and 
satellite subhalo is often not clear-cut, especially for halos undergoing rapid 
merger activity which are highly disturbed and aspherical.
In particular, our tracking also can produce ``switches'': if a satellite 
becomes more massive than the central, it becomes the central while the 
central becomes a satellite, often switching back in the next output.\footnote{
In more detail, a switch occurs when the density peak of a satellite (above the 
background) contains more mass than is within the central subhalo's radius at 
the position of the satellite}. 
In these cases, there is typically not a well-defined single peak that 
represents the center of the halo profile.

We assign to a subhalo its mass when it fell into its current host halo, i.e., 
its subhalo mass when it was last a central subhalo, $M_{\rm inf}$.
Centrals are assigned $M_{\rm inf}$ as their current bound mass.
Subhalo infall mass has been shown to correlate with galaxy stellar mass 
\citep{ValOst06,WanKauDeL06,YanMovdB08}.\footnote{
As has subhalo maximum circular velocity at infall, $V_{\rm c,inf}$ 
\citep{ConWecKra06,BerBulBar06}.
We compare $M_{\rm inf}$ and $V_{\rm c,inf}$ in detail in Paper I.}
If two satellite subhalos merge, their satellite child is given the sum of 
their infall masses.
During a switch, when a satellite becomes a central, it would acquire all halo 
mass not bound in other subhalos.
To avoid strong fluctuations in $M_{\rm inf}$, a central subhalo is assigned 
its halo's current self bound mass only if it was a central in the same halo 
in the previous output.
If a central was a satellite or a central in a different (smaller) halo in the 
previous output, it is assigned the sum of its parents' $M_{\rm inf}$.

We select subhalos with $M_{\rm inf} > 10^{12} h^{-1}M_\odot$ in the larger 
simulation and scale down to $M_{\rm inf} > 10^{11} h^{-1}M_\odot$ in the 
smaller, higher resolution simulation, by requiring consistency between the two 
simulations in their overlap regime.\footnote{
Using the Millennium simulation \citep{Spr05}, \citet{KitWhi08} require an 
analytic model for satellite infall times after subhalo disruption to match 
small-scale galaxy clustering at $z \sim 0$.
We do not expect this numerical disruption to significantly bias our results 
since our $100\,h^{-1}$Mpc simulation has higher mass and temporal resolution.}
Halos of mass $10^{11}~(10^{12}) h^{-1}M_\odot$ cross below $M_*$, the 
characteristic mass of collapse, at $z=1.5~(z=0.8)$, so we probe $M>M_*$ 
subhalos for much of the redshift range we consider.
We also expect our sample of $M_{\rm inf} > 10^{12} h^{-1}M_\odot$ subhalos to 
approximately correspond to $L \gtrsim L_*$ galaxies at the redshifts we 
examine (see, e.g., \citet{ConWec08} for halo-galaxy mass relation based on 
abundance matching, but note that they match stellar mass to halo mass, which 
is typically $10-15\%$ higher than subhalo infall mass).
Additionally, most massive galaxies are gas-rich (blue) at $z \gtrsim 1$  \citep{Coo07,Ger07,HopCoxKer08}, possessing enough gas to be actively star 
forming.
Thus, we anticipate that most, if not all, mergers we track have the capacity 
to drive galaxy activity such as starbursts and quasars.

\subsection{Merger Criteria} \label{sec:mergerdef}

We select a subhalo as a major merger (henceforth merger) if its two most 
massive parents, with $M_{\rm inf,2}\leq M_{\rm inf,1}$, satisfy 
$M_{\rm inf,2}/M_{\rm inf,1} > 1/3$.
As mentioned above, galaxy mergers with stellar mass ratios closer than 3:1 are 
expected to drive interesting activity, e.g.~quasars and starbursts.
Unless otherwise stated, we use the shortest simulation output spacing to 
define the merger time interval, corresponding to $50\,$Myr ($\sim 100\,$Myr) 
for $M_{\rm inf} > 10^{11}~(10^{12}) h^{-1}M_\odot$ at $z>2.5$, and 
$\sim 200\,$Myr for all masses at $z<1.6$.\footnote{
This gives $280$ ($490$) mergers at $z=2.6$ ($z=1$) above the lower mass cut in 
our smaller, higher resolution simulation.}

Other definitions of mergers produce significantly different merger samples.
In related work, which inspired our investigation, \citet{ThaScaCou06} used a 
dark matter plus hydrodynamic simulation to measure the small-scale clustering 
of subhalos with recent large mass gains, finding that these subhalos have 
enhanced small-scale clustering relative to a population with the same 
large-scale ($\gtrsim 1\,h^{-1}$Mpc) clustering.
Mass gain is convenient in that it does not require histories beyond the 
previous time step, and mass gain is unambiguously defined for all subhalos.
However, using our simulations and subhalo finder, the resulting sample is 
almost entirely different from the one defined above.\footnote{
For $M_{\rm inf}> 10^{11}\,h^{-1}M_\odot$ subhalos at $z=2.6$, $492$ have 
$M_{\rm cont,2}/M_{\rm cont,1} > \frac{1}{3}$, compared to the $260$ for our 
infall mass ratio definition, with only $7$ overlapping the two sets.}
Specifically, using a mass gain merger definition in our simulations led to 
`mergers' where the two contributing galaxies almost always remained as 
distinct entities after the merger event.
The most common instance of major mass gain is a satellite subhalo gaining mass 
during its movement within its host halo, particularly as it moves away from 
the halo center (see Fig.~2 in Paper I and \citet{DieKuhMad07} for examples).
A subhalo can also gain mass by stripping material from the outskirts of a 
nearby subhalo.
In $75\%$ of the cases of major mass gain, one of the progenitors contributed 
less than $10\%$ of its mass to the resulting `merged' child.
The most bound particles (where we expect the stellar component of a galaxy) 
were unaffected.
We did find significantly increased small-scale clustering for these mass gain
subhalos, attributable to the remaining nearby subhalo which just `merged' with 
it.
Similar issues in using mass gain to define mergers were noted in \citet{Mau07}.

\section{Small-Scale Spatial Clustering} \label{sec:xi}

\begin{figure*}
\begin{center}
\resizebox{3in}{!}{\includegraphics{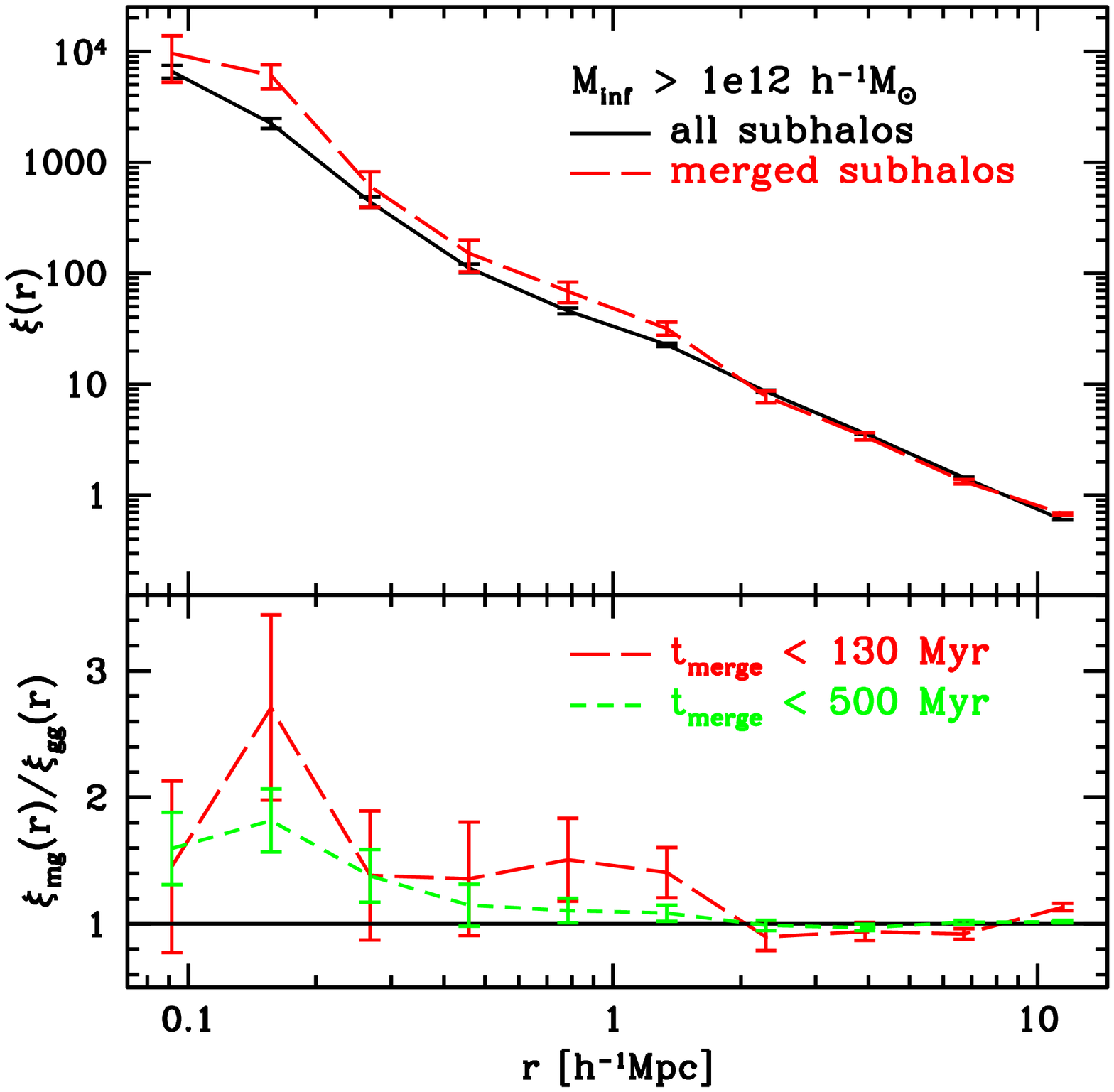}}
\resizebox{3in}{!}{\includegraphics{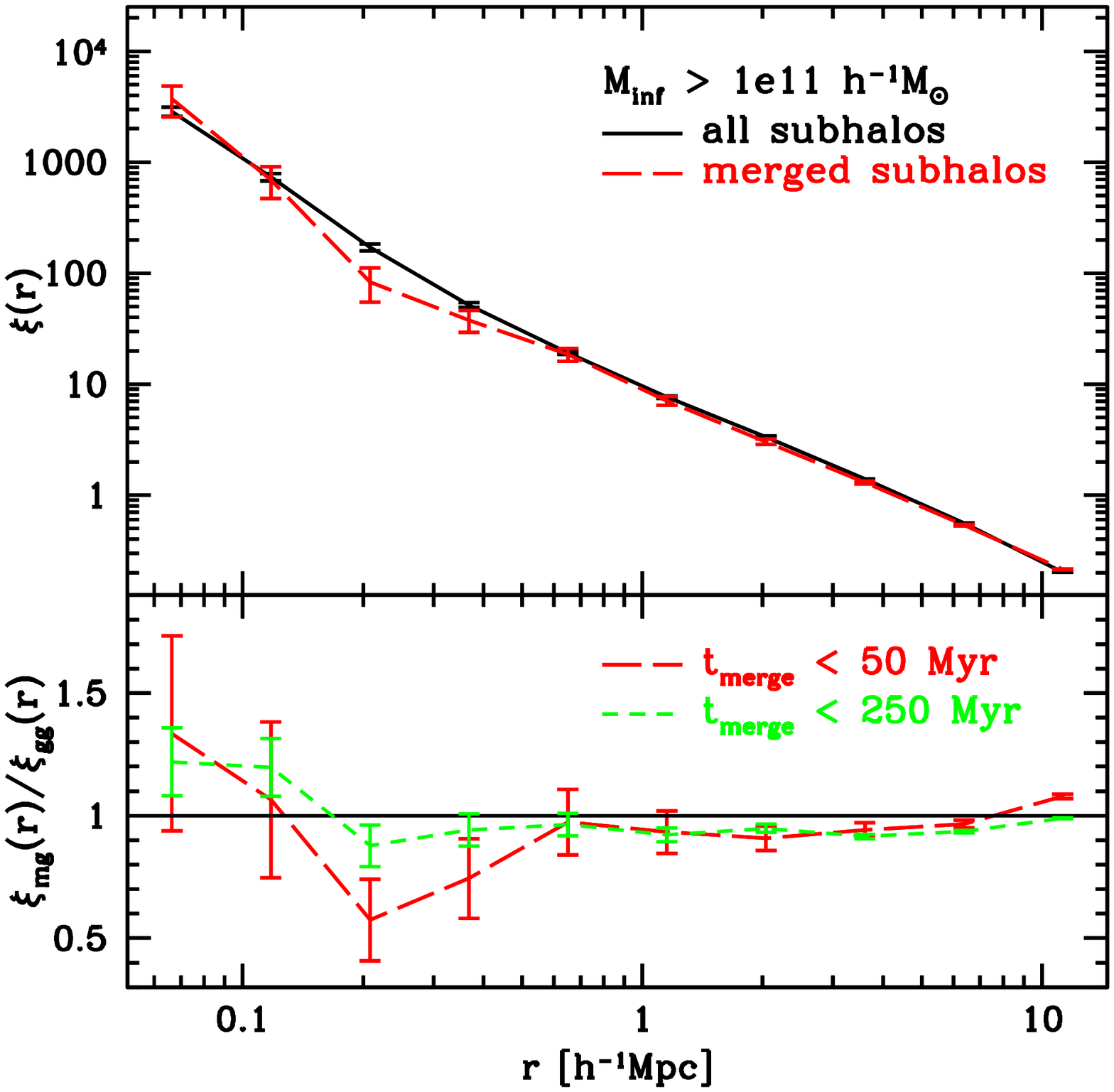}}
\end{center}
\vspace{-0.1in}
\caption{
Cross-correlation of recently merged subhalos with all subhalos, $\xi_{mg}(r)$, 
and the auto-correlation of all subhalos, $\xi_{gg}(r)$, at $z=2.6$, for 
$M_{\rm inf} > 10^{12} h^{-1}M_{\odot}$ (left) and
$M_{\rm inf} > 10^{11} h^{-1}M_{\odot}$ (right).
Infall mass, $M_{\rm inf}$, is matched between the merged and full subhalo 
samples.
\textbf{Top Left}: $\xi_{mg}(r)$ for subhalos merging within the last 
$130\,$Myr (long-dashed) and $\xi_{gg}(r)$ for all subhalos (solid).
\textbf{Bottom Left}: Ratio of the cross- and auto-correlations above, for 
mergers within the last $130\,$Myr (long-dashed) and $500\,$Myr (short-dashed).
\textbf{Top Right}: $\xi_{mg}(r)$ for subhalos merging within the last
$50\,$Myr (long-dashed) and $\xi_{gg}(r)$ for all subhalos (solid).
\textbf{Bottom Right}: Ratio of the cross- and auto-correlations above, for 
mergers within the last $50\,$Myr (long-dashed) and $250\,$Myr (short-dashed).
Higher mass subhalos show stronger enhanced clustering from mergers, but at 
both masses no signal persists for subhalos $>500\,$Myr after merging.
Errors are given by $\sqrt{N_{\rm pair}}$ and do not include sample variance.
} \label{fig:xi}
\end{figure*}

When examining the effects of recent mergers on spatial clustering it is
important to define an appropriate comparison sample.
We have chosen all subhalos above the same given infall mass cut as the 
mergers, with a matched $M_{\rm inf}$ distribution.
We match $M_{\rm inf}$ to remove any possible artificial biasing from, e.g., 
mergers preferentially occurring for subhalos of higher mass.
Using only a mass cut without matching the mass distribution leads to a similar, 
but weaker, effect.
If infall mass is a good proxy for stellar mass, our merger and comparison 
samples correspond to populations matched in stellar mass.

To measure the small-scale ($\sim 100-1000\,h^{-1}$kpc) clustering of subhalo 
mergers relative to the clustering of the general subhalo population, we 
measure the cross-correlation function\footnote{Not only is the
cross-correlation of the mergers with the general population interesting in 
itself, but it also provides better statistics.} of the merged and general
population, $\xi_{mg}$, and the auto-correlation function of the general
population, $\xi_{gg}$.
Our limited volume unfortunately does not allow us to sub-divide our simulation
to measure the sample variance error.
We show $\sqrt{N_{\rm pair}}$ errors on the correlation function points, but we 
caution that this may underestimate the error by up to a factor of 2.
Our clustering measurements are limited on small scales by the force resolution 
and on large scales by the simulation volume.
We present results on scales where these effects are minor, and we further 
discuss the effects of finite simulation volumes and the statistics of massive 
halos in \S\ref{sec:censathod}.

Figure \ref{fig:xi} shows the spatial clustering of recently merged subhalos 
and the general subhalo population at $z=2.6$, for two mass regimes and several 
merger time intervals.
These results are representative of our results at other redshifts.
Over the smallest time intervals, both high and low mass subhalos have a rise 
and a decline relative to the general population, with the rise being most 
prominent for higher mass subhalos and the decline most prominent for lower 
mass subhalos.
For $M_{\rm inf}>10^{11}\,h^{-1}M_\odot$, we find an upper limit of $1.8\times$ 
enhancement at $70\,h^{-1}$kpc, increasing to $3\times$ that of the general
population at $\sim150\,h^{-1}$kpc for $M_{\rm inf}>10^{12}\,h^{-1}M_\odot$.
In addition, lower mass subhalos exhibit a deficit at $100-300\,h^{-1}$kpc.
The enhancement/deficit declines rapidly with the time since the merger, and we
see no signal $>500\,$Myr after the merger.
As time progresses central mergers will become satellites in larger halos and
satellite mergers will move within their host halos (and perhaps merge with the 
central), washing out the correlation between the merger and its halo 
properties that we see at the time of the merger.

\section{Halo Occupation Distribution and Radial Profile} \label{sec:hod}

The halo model \citep{PeaSmi00,Sel00,CooShe02} can provide insight into the 
observed clustering signals of the merged subhalo population in 
Fig.~\ref{fig:xi}.
In this framework, galaxies populate dark matter halos such that their 
large-scale spatial clustering is determined primarily by the clustering of 
their host dark matter halos (``2-halo term''), while their small-scale spatial 
clustering arises from galaxies in the same host halo (``1-halo term'').
The objects occupy dark matter halos according to a Halo Occupation 
Distribution (HOD) and have some radial profile within these halos.
Since the clustering of recently merged galaxies differs from that of all 
galaxies, we expect mergers to differ from the general galaxy population in 
their HOD and/or profile.

\subsection{HOD of Subhalos}

\begin{figure*}
\begin{center}
\resizebox{3in}{!}{\includegraphics{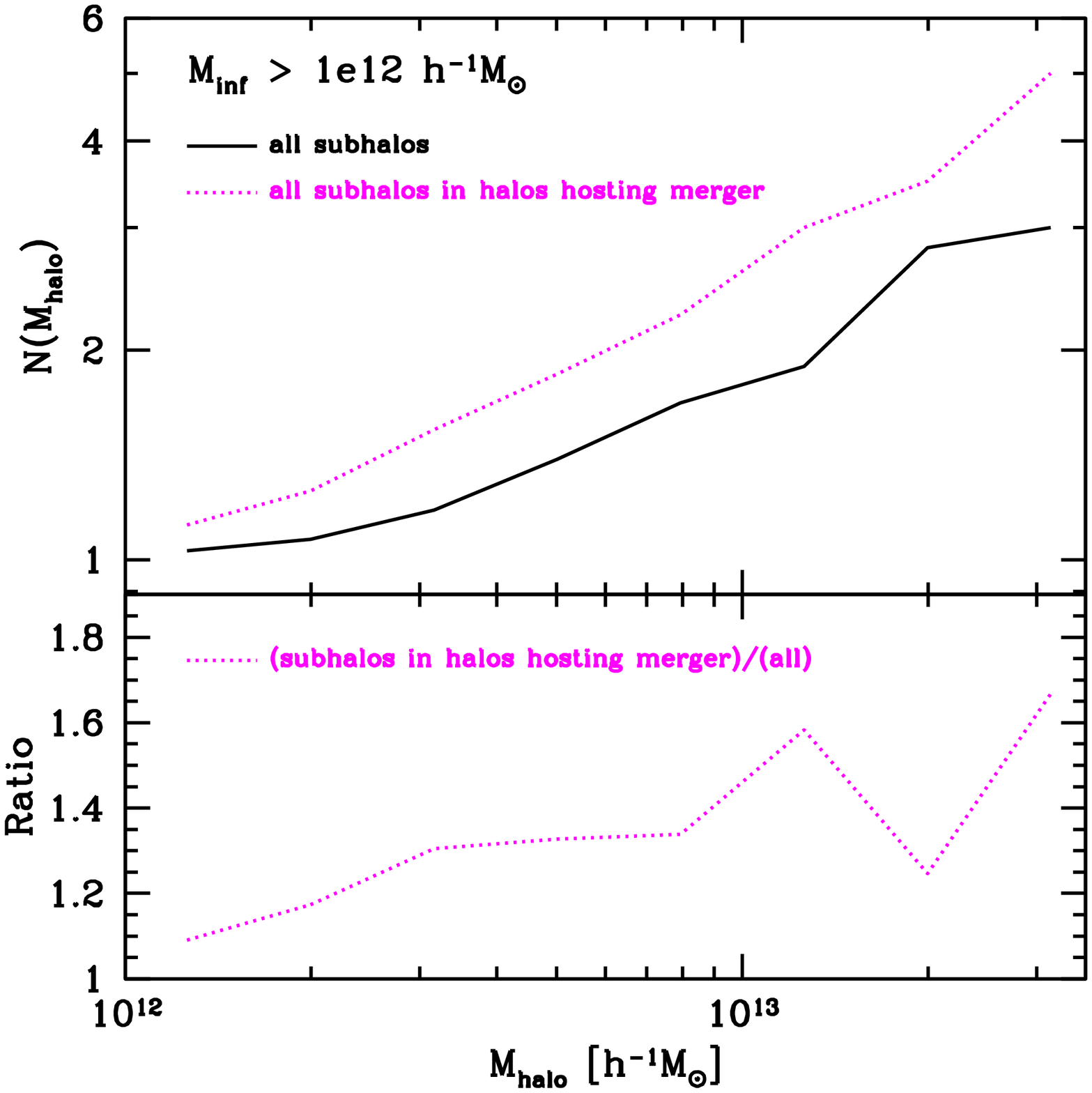}}
\resizebox{3in}{!}{\includegraphics{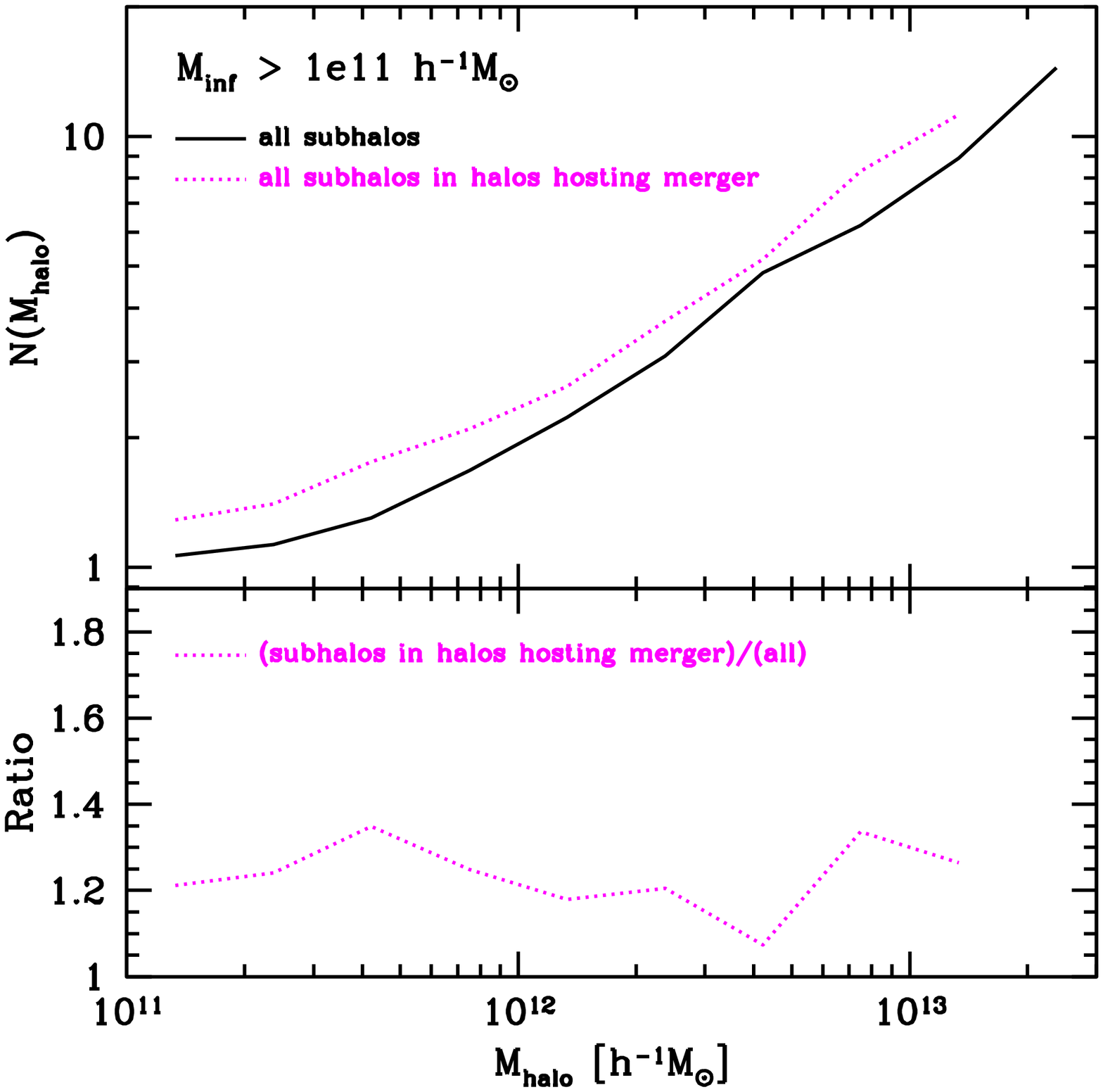}}
\end{center}
\vspace{-0.1in}
\caption{
\textbf{Top Panels}: Halo Occupation Distribution (HOD) at $z=2.6$ for all 
subhalos (solid) and subhalos in a halo hosting a recently merged subhalo 
(dotted).
Shown are subhalos with $M_{\rm inf}>10^{12}h^{-1}M_\odot$ and mergers occuring 
within $130\,$Myr (left), and $M_{\rm inf}>10^{11}h^{-1}M_\odot$ and mergers 
occurring within $50\,$Myr (right).
\textbf{Bottom Panels}: Ratio of the above HOD's of subhalos in a halo hosting 
a merger to that for all subhalos.
There are more subhalos in halos with recently merged subhalos, and the effect 
is stronger for more massive subhalos.
For recently merged massive subhalos, which show a strong increase in 
small-scale clustering, the relative subhalo excess in halos with mergers tends 
to increase with halo mass.
Similar trends persist at all redshifts we probe.
} \label{fig:hodsubhalos}
\end{figure*}

Figure \ref{fig:hodsubhalos} shows the HOD of subhalos (central and satellite)
at $z=2.6$ (corresponding to the populations in Fig.~\ref{fig:xi}) for all 
subhalos and for all subhalos within a halo containing a recently merged 
subhalo.
We see clearly that halos hosting subhalo mergers tend to have more subhalos.
This increase occurs for both our high and low mass samples and for all 
redshifts we probe.
For massive subhalo mergers the increase is larger and tends to rise to larger 
halo mass, which enhances the cross-correlation for mergers.
Although lower mass subhalo mergers also have more subhalos per halo, the 
relative number does not change strongly with increasing halo mass -- host halo 
mass does not significantly influence merger statistics.

Note that we expect some increase in the number of subhalos above a given mass 
threshold from mergers of subhalos just below the mass threshold, while mergers 
between subhalos above the threshold decrease the number of subhalos.
Which effect wins out requires detailed simulations such as ours.

\subsection{Central and Satellite Cross-Correlation} \label{sec:censat}

We now distinguish between contributions from satellite and central subhalos to 
the correlation function and HOD of mergers and the general population.
There are inherent subtleties in this breakdown; 
as mentioned earlier, the identification of satellite vs. central subhalos is 
not entirely clear-cut at these masses and redshifts.
In particular, one type can switch to another, and often does for mergers (see 
Paper I for more detail).
Also, other tracking schemes might alter the relation between central and 
satellite assignments we use.
If switched satellite mergers are assigned as centrals, this will of course not 
change the observed clustering signal but would alter where the contribution 
from our switches shows up in the breakdown of central and satellite effects.

\begin{figure}
\begin{center}
\resizebox{3in}{!}{\includegraphics{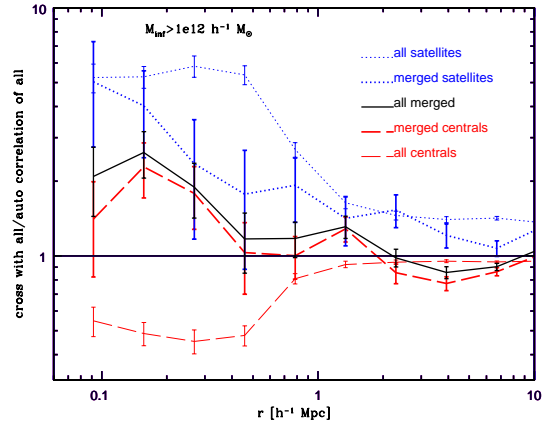}}
\end{center}
\vspace{-0.1in}
\caption{
The cross-correlation of the component populations (merged and all satellites, 
merged and all centrals) with all subhalos (the latter with $M_{\rm inf}$ 
matched distribution), at $z=2.6$ for $M_{\rm inf} > 10^{12} h^{-1}M_{\odot}$, 
corresponding to the left hand panel of Fig.~\ref{fig:xi}.
Top to bottom at far left are
$\xi_{\rm sat-all} /\xi_{\rm all}$,
$\xi_{\rm sat, merged-all} /\xi_{\rm all}$,
$\xi_{\rm all, merged-all} /\xi_{\rm all}$,
$\xi_{\rm cen,merged-all} /\xi_{\rm all}$,
$\xi_{\rm all-all} /\xi_{\rm all}$,
$\xi_{\rm cen-all} /\xi_{\rm all}$.
} \label{fig:xisplit}
\end{figure}

Figure \ref{fig:xisplit} shows the ratios of the cross-correlations for central 
mergers and satellite mergers to the auto-correlation of the full sample (the 
left hand side of Fig.~\ref{fig:xi} above), for 
$M_{\rm inf} > 10^{12} h^{-1}M_{\odot}$ subhalos at $z=2.6$.
The ratios of the cross-correlations of the full central and satellite 
populations to the full subhalo population are also shown.
As before, the full population is matched in $M_{\rm inf}$ to the merged 
population.
In general, satellites have a larger cross-correlation with all subhalos 
compared to centrals because the minimum host halo mass for a subhalo of a 
given mass is larger for a satellite than a central subhalo (increasing the 
large-scale clustering), and a satellite will always have a central in its halo 
but not vice versa (increasing the small-scale clustering).
The satellite and central cross-correlations must be summed, weighted by their 
population's fraction of the full population to get the full cross-correlation.

The merged central and merged satellite contributions to the cross-correlation 
differ from their counterparts in the full population.
Generally, the merged satellites have decreased small-scale clustering, 
relative to all satellites, while merged centrals have enhancement relative to 
all centrals.
In addition, satellites comprise a larger fraction of the merger population 
than of the full subhalo population (typically $\sim 1.5-2\times$ larger, see  
Paper I), i.e., 
$N_{\rm sat, merged}/N_{\rm cen, merged} \neq N_{\rm sat}/N_{\rm cen}$, 
changing the relative weights satellite and central contributions in the full
cross-correlation.\footnote{
Since satellites preferentially reside in higher mass halos, if the recently 
merged population is comprised of a higher fraction of satellites, it will also 
exhibit enhanced large-scale clustering.}
These three trends persist at all redshifts and subhalo masses, but their 
relative strengths vary to give the different behavior seen in 
Fig.~\ref{fig:xi} for different subhalo mass cuts.

\subsection{Central and Satellite HOD} \label{sec:censathod}

\begin{figure*}
\begin{center}
\resizebox{3in}{!}{\includegraphics{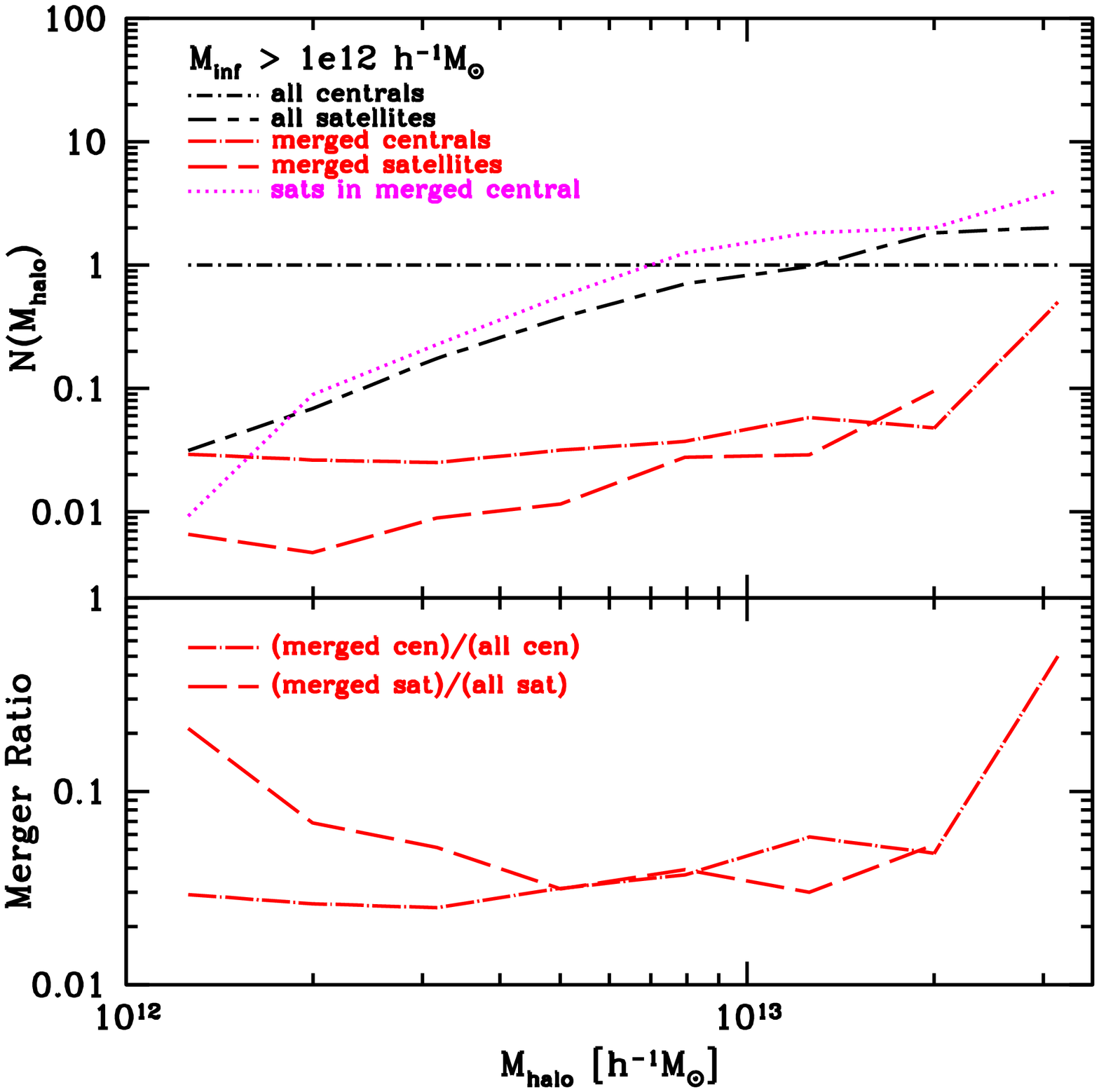}}
\resizebox{3in}{!}{\includegraphics{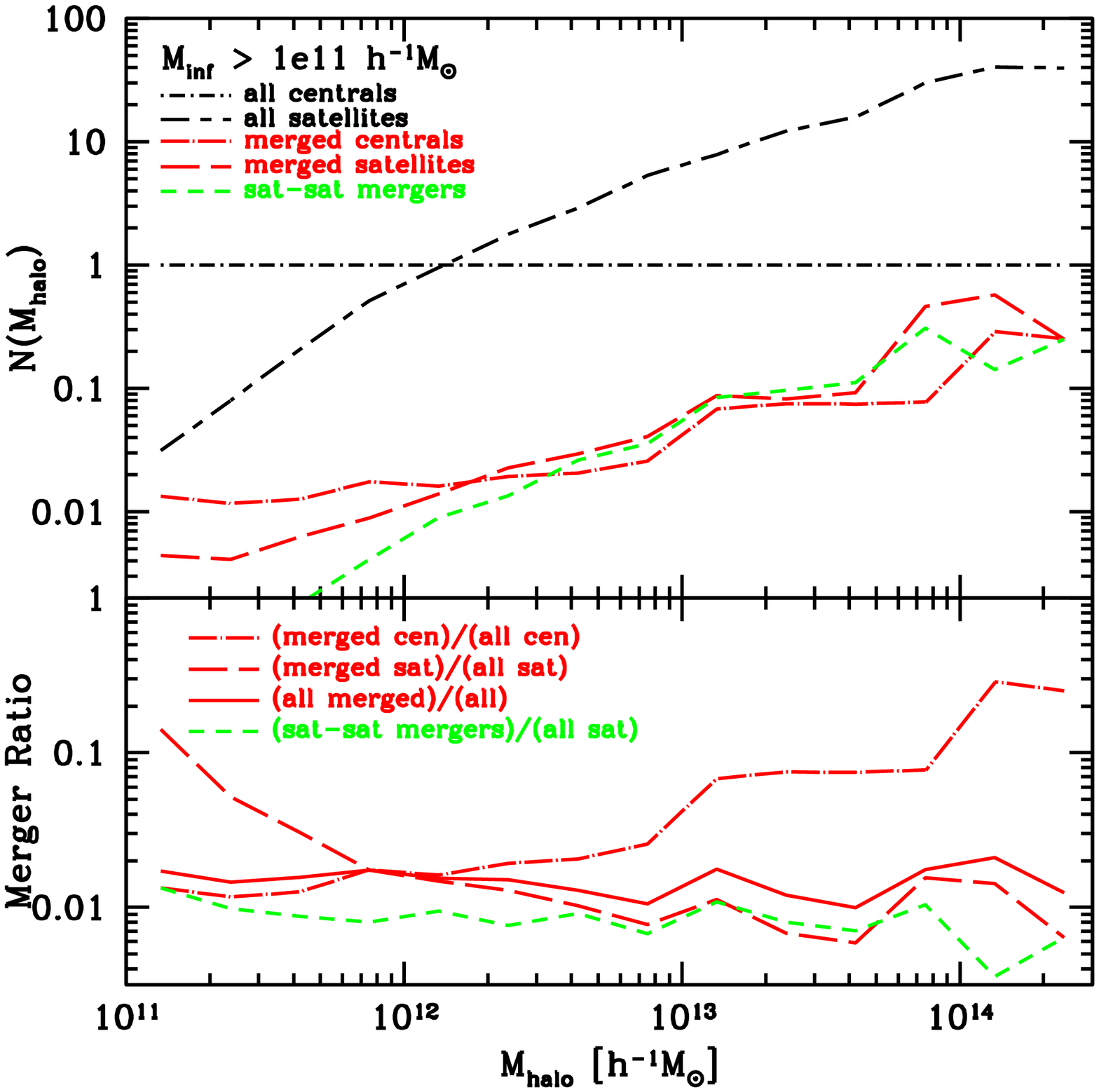}}
\end{center}
\vspace{-0.1in}
\caption{
Halo Occupation Distribution (HOD), broken down in terms of merger types: 
centrals, satellites, recently merged centrals, and recently merged satellites.
\textbf{Top Left}: Subhalos with $M_{\rm inf}>10^{12}h^{-1}M_\odot$ at $z=2.6$, 
corresponding to the left hand side of Fig.~\ref{fig:xi} and the component 
contributions in Fig.~\ref{fig:xisplit}.
The merger time interval is $130\,$Myr.
Also shown is the HOD for satellites in halos with a recently merged central.
\textbf{Bottom Left}: Ratio of HOD's of recently merged centrals to all 
centrals and of recently merged satellites to all satellites.
\textbf{Top Right}: Subhalos with $M_{\rm inf}>10^{11}h^{-1}M_\odot$ at $z=1$
and mergers occurring within $230\,$Myr, with cross-correlation similar to the 
$250\,$Myr interval in Fig.~\ref{fig:xi} right at $z=2.6$.
The HOD is also shown for satellite-satellite mergers.
No strong excess is seen for satellites in halos with a recently merged central 
(not shown).
\textbf{Bottom Right}: Ratio of HOD's of recently merged centrals to all 
centrals, recently merged satellites to all satellites, and satellite-satellite 
mergers to all satellites.
Also shown is the ratio of HOD's of subhalo mergers to all subhalos (regardless 
of type).
Satellite-satellite mergers, and subhalo mergers (regardless of type), show no 
strong dependence on halo mass, though our statistics are not sufficient to 
rule out a cutoff at the highest mass.
} \label{fig:hod}
\end{figure*}

We can employ this central/satellite split for the HODs as well.
Figure~\ref{fig:hod} shows the HODs for two mass ranges and two redshifts: 
at left is the HOD for high-mass subhalo mergers at $z=2.6$, corresponding to 
Fig.~\ref{fig:xi} (left), and at right is the HOD corresponding to the less 
massive subhalos at $z=1$, for a longer ($230\,$Myr) merger time interval.
The cross/auto-correlation ratio for this latter sample is similar to that 
shown in Fig.~\ref{fig:xi} (right), but we show the results at lower redshift 
to illustrate the trends we see over a larger host halo mass range.

Figure~\ref{fig:hod} (left) shows that {\it high-mass\/} recently merged central 
subhalos occupy halos with an excess of satellites (relative to the average) at 
most host halo masses (dotted curve).
The physical extent of the enhancement, $\sim300\,h^{-1}$kpc, coincides with 
the virial radius of the most massive halos 
($\sim 4\times 10^{13}\,h^{-1}M_\odot$) in our simulations at $z\sim2.6$.
The enhancement of the satellite population in halos hosting recent central 
mergers is not as large for lower mass subhalos (the curve for this quantity is 
almost indistinguishable from that of the full satellite population, thus it 
is not shown in Fig.~\ref{fig:hod} right).
This corresponds to the weakened enhancement, and decrement, in 
Fig.~\ref{fig:xi}.
In terms of the central/satellite subhalo breakdown, the increased satellite 
number for recently merged central subhalos has the strongest correspondence 
with increased small-scale clustering.

The other trends that we see appear across our redshift and mass regimes.
For a fixed subhalo mass cut, the fraction of centrals that have experienced a 
recent merger increases with halo mass.
This causes an enhancement of the central merger cross-correlation with the 
full population, since the rise in central merger fraction with halo mass gives 
the overall recently merged central population proportionally more satellites 
when all halo masses are summed.

It is worth noting two points about the relative increase of central mergers in 
higher mass halos.
The increase itself might be surprising, as both the number of satellites and 
the dynamical friction timescale increase approximately linearly with halo mass.
One expects that the increase of possible satellites and the slowdown of their 
approach to the center would then cancel, rather than producing increased 
numbers of central mergers.
Looking instead at merger parent types, we found that the central-satellite 
merger HOD does not increase as steeply as the merged central HOD.
Central mergers in the highest mass halos are often satellite-satellite mergers 
whose child is a central (switches), rather than central-satellite mergers.
While this is a small fraction ($3\%$) of all central mergers, it is a larger 
fraction of those in high-mass halos.

Secondly, the increase of central mergers to higher mass halos suggests a 
possible error from neglecting high mass halos which are too rare to occur in 
the simulation volume.
Their effect can be estimated analytically by extrapolating the central merger 
fraction as a function of halo mass, using the halo model to find the relative 
contribution of merged and all centrals to the cross-correlation at some given 
radius, and seeing how this changes as larger masses are included.
We find that, on average, neglecting halos more massive than those found in 
our simulations only causes a small change in the cross-correlation of centrals 
with the full population.
However, our largest halos are quite rare, and so the merger fraction in them 
does not always tend to the average.
The rise in central merger fraction with halo mass is largest for our 
$M_{\rm inf}>10^{11} h^{-1} M_\odot$ subhalos and thus of most concern.
As a check, we averaged our cross-correlation for the smaller simulation over 
several outputs to confirm the trends we see in Fig.~\ref{fig:xi} (right).
The output to output variation does average to this trend, but individual 
outputs can show different sign effects, albeit with significant error bars.
This is especially true at lower redshift, where our imperfect sampling of high 
mass halos plays a larger role.
It might be counterintuitive that $300\,h^{-1}$kpc clustering is not very well 
measured in a $100\,h^{-1}$Mpc box, but the peak of the power spectrum is at 
large scales in a $\Lambda$CDM model and the scatter in the cross-correlation 
is driven by the contribution from rare, massive halos which host many 
satellites.

We now turn to the satellite HOD.
Figure~\ref{fig:hod} (bottom) shows that the fraction of merged satellites to 
all satellites decreases with increasing host halo mass (long-dashed curve).
However, a significant fraction of merged satellites are satellite-central 
switches, which dominate in low-mass halos where a central and satellite are 
more likely to be comparable in mass and thus able to switch.
Additionally, some merged satellites are central-satellite mergers which then 
fell into the host halo.
Examining instead merger parent type, satellite-satellite mergers are 
essentially a constant fraction of satellite subhalos across all halo masses 
(short-dashed line).
We see a slight increase for massive host halos at $z=2.6$ and no increase at 
$z=1$.
Additionally, we see weak evidence for a cutoff in the satellite-satellite 
merger HOD for the most massive halos, consistent with the idea that increasing 
relative velocities of satellites in the most massive halos cuts off 
satellite-satellite mergers \citep{MakHut97}.
However, the high-mass halo statistics are poor in our modest simulation 
volumes.

A simple argument can be made which reproduces the observed scaling of the 
number of satellite-satellite mergers with the overall satellite population, 
based upon the HOD scaling with increasing halo mass/volume.
For satellites with random orbital parameters in a halo of a fixed mass, the 
satellite-satellite merger probability will scale as number density squared 
($n_{\rm sat}^2$), so the fraction of satellites undergoing satellite-satellite 
mergers scales as $n_{\rm sat}$.
For halos above a few times the mass of a given satellite population, the 
number of satellites scales with the halo mass \citep[e.g.,][]{KraBerWec04}, 
which scales approximately with the halo volume, so 
$N_{\rm sat} \propto M_{\rm halo} \propto V_{\rm halo}$.
Thus, $n_{\rm sat} \approx$ constant, so the satellite-satellite merger fraction 
remains roughly constant.

Finally, the lower right panel of Fig.~\ref{fig:hod} shows that when we do not 
split by type, the subhalo merger fraction is essentially independent of halo 
mass.
Again, this contradicts the frequently expressed intuition that mergers are 
less frequent in higher mass halos, though we caution that our statistics are 
poor for cluster-mass halos, and we are working at high redshift.

\subsection{Radial Distribution Profile} \label{sec:profile}

\begin{figure}
\begin{center}
\resizebox{3in}{!}{\includegraphics{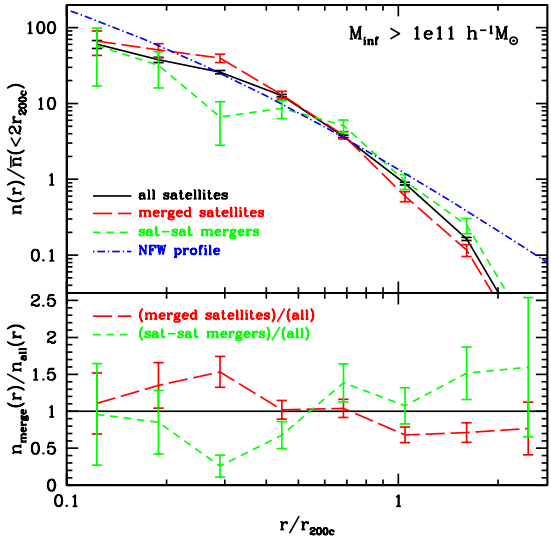}}
\end{center}
\vspace{-0.1in}
\caption{ 
\textbf{Top}: Halo radial distribution profile of satellites with 
$M_{\rm inf} > 10^{11} h^{-1}M_\odot$ at $z=2.6$, for all satellites (solid), 
recent mergers (long-dashed), and mergers from satellite-satellite parents 
(short-dashed).
The merger time interval is $50\,$Myr.
Also shown is the halo NFW density profile (dot-dashed), using an NFW 
concentration of $c=4$, typical for halos at this mass and redshift.
Density distributions are normalized to the average density of the population 
within $2r_{200c}$.
\textbf{Bottom}: Ratio of the recently merged satellite normalized density to 
that of all satellites (long-dashed) and the same for recent mergers from 
satellite-satellite parents (short-dashed).
We find no dependence of these results on satellite infall mass or redshift.
} \label{fig:profile}
\end{figure}

Figure \ref{fig:profile} shows the other ingredient required to predict 
clustering: the (stacked) satellite radial distribution profile, for subhalos 
with $M_{\rm inf}>10^{11}\,h^{-1}M_\odot$ at $z=2.6$ and a merger time interval 
of $50\,$Myr.
We choose this mass regime and redshift because the high temporal resolution 
allows us to accurately identify the locations of mergers, though the behavior 
across all our mass and redshift regimes is consistent with these results.
Shown are the profiles for all satellites, recently merged satellites, 
recently merged satellites from satellite-satellite parents (because of 
switches), and the NFW \citep{NFW96} halo density profile.
We normalize the density distributions to the average value within 
$2r_{200c}$.\footnote{
We calculate the halo virial radius, $r_{200c}$, i.e. the radius within which 
the average density is $200\times$ the critical density, from the FoF 
($b=0.168$) mass by first converting to $M_{200c}$ assuming a spherical NFW 
density profile, and then taking 
$M_{200c} = 200 \frac{4\pi}{3} \rho_c r^3_{200c}$.}
We use $2r_{200c}$ because an appreciable fraction of satellite mergers occur 
just outside $r_{200c}$ given the highly aspherical geometry of halos at this 
mass and redshift regime.
The satellite subhalo profile well-traces the NFW profile from $r_{200c}$ down 
to $0.2r_{200c}$, and shows no dependence on subhalo mass, consistent with the 
distribution of subhalos selected on infall mass at $z=0$ of 
\citet{NagKra05}.\footnote{
When selecting subhalos on instantaneous bound mass, more massive halos are 
preferentially biased to the outer regions of a halo, since mass stripping 
leaves few massive subhalos in the inner region of a halo 
\citep[see][]{NagKra05}.}
The deficit below $0.2r_{200c}$ is a result of finite spatial resolution, tidal 
disruption of subhalos near the halo center, and subhalos temporarily 
disappearing from the sample during a fly-by (see Fig.~1 of Paper I).

As Fig.~\ref{fig:profile} shows, the radial distribution profile of satellite 
mergers approximately follows that of the entire satellite population.
The profile can be thought of as a measure of how satellite mergers correlate 
with centrals, bearing in mind that the distances are scaled by the halo virial 
radius.
Relative to the density distribution of all satellites, recently merged 
satellites have a slightly more concentrated profile, with enhanced probability 
of being at $20\%-40\%$ of the virial radius (long-dashed).
This produces a relative deficit in clustering of recently merged satellites 
out to the halo virial radius.
Considering instead recently merged centrals, we find that the profile of 
satellites in halos with a recently merged central traces the general satellite 
population without significant deviation (not shown).

While recently merged satellites exhibit a more concentrated profile, mergers 
between satellite-satellite subhalos preferentially occur in halo outskirts and 
are comparatively less common in the central regions (short-dashed curve).
Thus, the enhanced probability of finding recent satellite mergers at small 
scales is driven by switches (when a satellite merges with a central and 
results in a satellite).

\section{Merger Pairs} \label{sec:pairs}

While we have focused on measuring merger clustering via the cross-correlation 
of mergers with the general population, we can also measure directly the 
statistics of our limited number of merger pairs and distinguish the different 
origins for the pair members.
We focus on pairs of subhalos within $250\,h^{-1}$kpc which have both undergone 
a merger within the last $\sim250\,$Myr.
In all such cases of close merger pairs, both subhalos inhabit the same halo.
These pairs are quite rare: at $z=2.6$, there are $24$ ($2$) per 
($100\,h^{-1}$Mpc)$^3$ for $M_{\rm inf}>10^{11} (10^{12})\,h^{-1}M_{\odot}$, 
while at $z=1$ there are $2$ per ($100\,h^{-1}$Mpc)$^3$ for 
$M_{\rm inf}>10^{11}\,h^{-1}M_{\odot}$ and none at the higher mass.

For two subhalos within the same halo, the two types of possible pairings are 
satellite-central and satellite-satellite.
For $M_{\rm inf} > 10^{12}\,h^{-1}M_{\odot}$, all pairs are composed of 
satellite-central subhalos, while for $M_{\rm inf} > 10^{11}\,h^{-1}M_{\odot}$, 
$25\%$ of close merger pairs are satellite-satellite subhalos.
In all cases, the central-satellite pairs arise when a satellite-central 
merger occurs simultaneously with a satellite-satellite merger in a single halo.
For the rarer cases of satellite-satellite merger pairs, two-thirds of the 
recently merged satellites arise from satellite-satellite parents within a 
halo, and one third arise from a satellite-central merger (a switch).

\section{Summary and Discussion} \label{sec:discussion}

Using high-resolution dark matter simulations in cosmological volumes, we have 
measured the small-scale spatial clustering for massive subhalo mergers at high 
redshift and compared against the clustering of the general subhalo population 
of the same mass.
We have described the merger populations in terms of their HOD and radial 
profile, including the contributions of centrals vs. satellites.
We assign subhalos their mass at infall, as a proxy for galaxy stellar mass, 
but make no further attempt to model the baryonic component.
We consider mergers with $<3$:1 infall mass ratios, motivated by the 
expectation that these can trigger activity such as quasars or starbursts.
Our main results are:

\begin{itemize}
\item At $z=2.6$, recently merged, massive 
($M_{\rm inf}>10^{12}\,h^{-1}M_\odot$) subhalos exhibit enhanced small-scale 
clustering compared to random subhalos with the same infall mass distribution.
This excess peaks at $100-300\,h^{-1}$kpc, while the clustering exhibits a dip
at slightly larger scales. 
Lower mass subhalos ($M_{\rm inf}>10^{11}\,h^{-1}M_\odot$) exhibit signs of a 
small rise in clustering at $<100\,h^{-1}$kpc, with a deficit at slightly 
larger scales, though our results are noisy in this regime.
We find similar behavior at $z=1$.
The merger signal weakens rapidly with time, vanishing for time intervals 
longer than $500\,$Myr after the merger.
\item Considering their HOD, halos hosting recently merged subhalos tend to 
have more subhalos.
This enhancement is stronger for more massive subhalos and exhibited growth 
with halo mass.
\item Breaking the contributions to the cross-correlation into those from 
satellite and central subhalos, the recently merged massive centrals, which  
show the largest enhancement of small-scale clustering, preferentially occupy 
halos with more satellites.
More generally, recently merged centrals occupy higher mass halos and recently 
merged satellites occupy lower mass halos with fewer (perhaps no other) 
satellites.
The resulting increase for the central cross-correlation, decrease for the 
satellite cross-correlation, and change in the ratio of their contributions 
combine to produce the cross clustering enhancement or decrement.
\item For the range of halo masses we probe, the (type-independent) subhalo 
merger fraction is independent of host halo mass.
We find similar behavior for the halo mass dependence of satellite-satellite 
mergers, i.e. the fraction of satellites that experience a merger with another 
satellite does not depend on halo mass, except perhaps in our poorly sampled 
largest mass halos ($\sim 10^{14}\,h^{-1} M_\odot$).
\item The radial profile of recently merged satellites roughly follows that of 
the entire satellite population, which also follows the halo density profile 
out to the virial radius.
Satellite-satellite mergers preferentially occur in the outer regions of a halo.
\item Mergers defined via major mass gain exhibit a strong small-scale 
clustering enhancement because significant mass gain is caused by interactions 
with a neighboring subhalo.
However, in these cases the neighbor exists both before and after the `merger' 
(i.e., no coalescence), and the major mass gain criterion does not correspond 
to dynamically disturbed subhalos.
Thus, we do not expect these objects to correlate with active galaxies.
\end{itemize}

Our measured subhalo merger enhancement suggests that, for populations with the 
same stellar mass at high redshifts, recent galaxy mergers should exhibit 
excess clustering at small radius, with a possible decrement in clustering at 
slightly larger scales.
This effect will persist for only a short period of time after the merger, 
with a stronger excess signal for galaxies of higher stellar mass.
Measuring this small-scale cross-correlation (against a general population) 
requires a surprisingly large volume ($\gtrsim 100\,h^{-1}$Mpc) because massive 
(satellite rich) halos contribute significant signal to the cross-correlation.
Thus, a fair sample of rich halos is necessary for robust conclusions.

A comparison of the clustering of our mergers to observational phenomena is a 
nontrivial future step, since many merger observables depend on complex gas 
physics.
Our results do not include whether the subhalos are gas-rich or not, though at 
the high redshifts we examine, we expect almost all galaxies to be gas-rich.
Satellites can be stripped of much of their gas during infall 
\citep[e.g.,][]{Dol08,Sar08}, though we have considered relatively massive 
satellites which have short infall times and thus experience less gas stripping.

Time scales for observables after a merger also have large uncertainty and 
scatter.
For example, an optical quasar might appear only $\sim 1\,$Gyr after the merger 
\citep[e.g.,][]{HopHerCox08,SprDiMHer05b}, by which time the enhanced 
clustering that we see has disappeared.
X-ray signals might appear sooner.\footnote{
We thank F. Shankar for suggesting this.}
Starbursts or starburst-related objects (e.g., sub-millimeter galaxies, Lyman 
break galaxies, ULIRGs) might also commence more rapidly after a merger 
\citep[e.g.,][]{Cox08} and thus appear a more promising analog to the effects 
we find here.

While we were preparing this work for preparation \citet{Ang08} appeared which 
also considers the host halos of satellite mergers, in the Millennium 
simulation.

\section*{Acknowledgments}

We thank the participants of the CCAPP workshop on AGN and cosmology for 
discussions, and especially Phil Hopkins, Alexie Leauthaud, Evan Scannapieco, 
Francesco Shankar, and Jeremy Tinker, and the referee for helpful comments and 
suggestions.
We also thank the CCAPP of The Ohio State University for hospitality in fall 
2007.
A.W. gratefully acknowledges the support of an NSF Graduate Fellowship, J.D.C. 
support from NSF-AST-0810820 and DOE, and M.W. support from NASA and the DOE. 
The simulations were analyzed at the National Energy Research Scientific 
Computing Center.

\bibliography{ms.bbl}

\begin{thebibliography}{}

\bibitem[\protect\citeauthoryear{{Angulo}, {Lacey}, {Baugh} \&
  {Frenk}}{{Angulo} et~al.}{2008}]{Ang08}
{Angulo} R.~E.,  {Lacey} C.~G.,  {Baugh} C.~M.,    {Frenk} C.~S.,  2008, ArXiv
  e-prints 0810.2177

\bibitem[\protect\citeauthoryear{{Barnes} \& {Hernquist}}{{Barnes} \&
  {Hernquist}}{1991}]{BarHer91}
{Barnes} J.~E.,  {Hernquist} L.~E.,  1991, \apjl, 370, L65

\bibitem[\protect\citeauthoryear{{Berrier}, {Bullock}, {Barton}, {Guenther},
  {Zentner} \& {Wechsler}}{{Berrier} et~al.}{2006}]{BerBulBar06}
{Berrier} J.~C.,  {Bullock} J.~S.,  {Barton} E.~J.,  {Guenther} H.~D.,
  {Zentner} A.~R.,    {Wechsler} R.~H.,  2006, \apj, 652, 56

\bibitem[\protect\citeauthoryear{{Blain}, {Chapman}, {Smail} \&
  {Ivison}}{{Blain} et~al.}{2004}]{Bla04}
{Blain} A.~W.,  {Chapman} S.~C.,  {Smail} I.,    {Ivison} R.,  2004, \apj, 611,
  725

\bibitem[\protect\citeauthoryear{{Blain}, {Smail}, {Ivison}, {Kneib} \&
  {Frayer}}{{Blain} et~al.}{2002}]{Bla02}
{Blain} A.~W.,  {Smail} I.,  {Ivison} R.~J.,  {Kneib} J.-P.,    {Frayer} D.~T.,
   2002, \physrep, 369, 111

\bibitem[\protect\citeauthoryear{{Bower}, {Benson}, {Malbon}, {Helly}, {Frenk},
  {Baugh}, {Cole} \& {Lacey}}{{Bower} et~al.}{2006}]{Bow06}
{Bower} R.~G.,  {Benson} A.~J.,  {Malbon} R.,  {Helly} J.~C.,  {Frenk} C.~S.,
  {Baugh} C.~M.,  {Cole} S.,    {Lacey} C.~G.,  2006, \mnras, 370, 645

\bibitem[\protect\citeauthoryear{{Carlberg}}{{Carlberg}}{1990}]{Car90}
{Carlberg} R.~G.,  1990, \apj, 350, 505

\bibitem[\protect\citeauthoryear{{Coil}, {Hennawi}, {Newman}, {Cooper} \&
  {Davis}}{{Coil} et~al.}{2007}]{Coi07}
{Coil} A.~L.,  {Hennawi} J.~F.,  {Newman} J.~A.,  {Cooper} M.~C.,    {Davis}
  M.,  2007, \apj, 654, 115

\bibitem[\protect\citeauthoryear{{Conroy} \& {Wechsler}}{{Conroy} \&
  {Wechsler}}{2008}]{ConWec08}
{Conroy} C.,  {Wechsler} R.~H.,  2008, arXiv: 0805.3346

\bibitem[\protect\citeauthoryear{{Conroy}, {Wechsler} \& {Kravtsov}}{{Conroy}
  et~al.}{2006}]{ConWecKra06}
{Conroy} C.,  {Wechsler} R.~H.,    {Kravtsov} A.~V.,  2006, \apj, 647, 201

\bibitem[\protect\citeauthoryear{{Cooper}, {Newman}, {Coil}, {Croton}, {Gerke},
  {Yan}, {Davis}, {Faber}, {Guhathakurta}, {Koo}, {Weiner} \&
  {Willmer}}{{Cooper} et~al.}{2007}]{Coo07}
{Cooper} M.~C.,  {Newman} J.~A.,  {Coil} A.~L.,  {Croton} D.~J.,  {Gerke}
  B.~F.,  {Yan} R.,  {Davis} M.,  {Faber} S.~M.,  {Guhathakurta} P.,  {Koo}
  D.~C.,  {Weiner} B.~J.,    {Willmer} C.~N.~A.,  2007, \mnras, 376, 1445

\bibitem[\protect\citeauthoryear{{Cooray} \& {Ouchi}}{{Cooray} \&
  {Ouchi}}{2006}]{CooOuc06}
{Cooray} A.,  {Ouchi} M.,  2006, \mnras, 369, 1869

\bibitem[\protect\citeauthoryear{{Cooray} \& {Sheth}}{{Cooray} \&
  {Sheth}}{2002}]{CooShe02}
{Cooray} A.,  {Sheth} R.,  2002, \physrep, 372, 1

\bibitem[\protect\citeauthoryear{{Cox}, {Jonsson}, {Somerville}, {Primack} \&
  {Dekel}}{{Cox} et~al.}{2008}]{Cox08}
{Cox} T.~J.,  {Jonsson} P.,  {Somerville} R.~S.,  {Primack} J.~R.,    {Dekel}
  A.,  2008, \mnras, 384, 386

\bibitem[\protect\citeauthoryear{{Croom}, {Boyle}, {Shanks}, {Smith}, {Miller},
  {Outram}, {Loaring}, {Hoyle} \& {da {\^A}ngela}}{{Croom}
  et~al.}{2005}]{Cro05}
{Croom} S.~M.,  {Boyle} B.~J.,  {Shanks} T.,  {Smith} R.~J.,  {Miller} L.,
  {Outram} P.~J.,  {Loaring} N.~S.,  {Hoyle} F.,    {da {\^A}ngela} J.,  2005,
  \mnras, 356, 415

\bibitem[\protect\citeauthoryear{{Croton}, {Gao} \& {White}}{{Croton}
  et~al.}{2007}]{CroGaoWhi06}
{Croton} D.~J.,  {Gao} L.,    {White} S.~D.~M.,  2007, \mnras, 374, 1303

\bibitem[\protect\citeauthoryear{{Davis}, {Efstathiou}, {Frenk} \&
  {White}}{{Davis} et~al.}{1985}]{DEFW}
{Davis} M.,  {Efstathiou} G.,  {Frenk} C.~S.,    {White} S.~D.~M.,  1985, \apj,
  292, 371

\bibitem[\protect\citeauthoryear{{Diemand}, {Kuhlen} \& {Madau}}{{Diemand}
  et~al.}{2007}]{DieKuhMad07}
{Diemand} J.,  {Kuhlen} M.,    {Madau} P.,  2007, \apj, 667, 859

\bibitem[\protect\citeauthoryear{{Dolag}, {Borgani}, {Murante} \&
  {Springel}}{{Dolag} et~al.}{2008}]{Dol08}
{Dolag} K.,  {Borgani} S.,  {Murante} G.,    {Springel} V.,  2008, arXiv:
  0808.3401

\bibitem[\protect\citeauthoryear{{Francke}, {Gawiser}, {Lira}, {Treister},
  {Virani}, {Cardamone}, {Urry}, {van Dokkum} \& {Quadri}}{{Francke}
  et~al.}{2008}]{Fra08}
{Francke} H.,  {Gawiser} E.,  {Lira} P.,  {Treister} E.,  {Virani} S.,
  {Cardamone} C.,  {Urry} C.~M.,  {van Dokkum} P.,    {Quadri} R.,  2008,
  \apjl, 673, L13

\bibitem[\protect\citeauthoryear{{Furlanetto} \& {Kamionkowski}}{{Furlanetto}
  \& {Kamionkowski}}{2006}]{FurKam06}
{Furlanetto} S.~R.,  {Kamionkowski} M.,  2006, \mnras, 366, 529

\bibitem[\protect\citeauthoryear{{Gao}, {Springel} \& {White}}{{Gao}
  et~al.}{2005}]{GaoSprWhi05}
{Gao} L.,  {Springel} V.,    {White} S.~D.~M.,  2005, \mnras, 363, L66

\bibitem[\protect\citeauthoryear{{Gawiser}, {Francke}, {Lai}, {Schawinski},
  {Gronwall}, {Ciardullo}, {Quadri}, {Orsi}, {Barrientos} et~al.,}{{Gawiser}
  et~al.}{2007}]{Gaw07}
{Gawiser} E.,  {Francke} H.,  {Lai} K.,  {Schawinski} K.,  {Gronwall} C.,
  {Ciardullo} R.,  {Quadri} R.,  {Orsi} A.,  {Barrientos} L.~F.,    et~al.,
  2007, \apj, 671, 278

\bibitem[\protect\citeauthoryear{{Gerke}, {Newman}, {Faber}, {Cooper},
  {Croton}, {Davis}, {Willmer}, {Yan}, {Coil}, {Guhathakurta} P.~and{Koo} \&
  {Weiner}}{{Gerke} et~al.}{2007}]{Ger07}
{Gerke} B.~F.,  {Newman} J.~A.,  {Faber} S.~M.,  {Cooper} M.~C.,  {Croton}
  D.~J.,  {Davis} M.,  {Willmer} C.~N.~A.,  {Yan} R.,  {Coil} A.~L.,
  {Guhathakurta} P.~and{Koo} D.~C.,    {Weiner} B.~J.,  2007, \mnras, 376, 1425

\bibitem[\protect\citeauthoryear{{Ghigna}, {Moore}, {Governato}, {Lake},
  {Quinn} \& {Stadel}}{{Ghigna} et~al.}{1998}]{Ghi98}
{Ghigna} S.,  {Moore} B.,  {Governato} F.,  {Lake} G.,  {Quinn} T.,    {Stadel}
  J.,  1998, \mnras, 300, 146

\bibitem[\protect\citeauthoryear{{Giavalisco}}{{Giavalisco}}{2002}]{Gia02}
{Giavalisco} M.,  2002, \araa, 40, 579

\bibitem[\protect\citeauthoryear{{Giavalisco}, {Steidel}, {Adelberger},
  {Dickinson}, {Pettini} \& {Kellogg}}{{Giavalisco} et~al.}{1998}]{Gia98}
{Giavalisco} M.,  {Steidel} C.~C.,  {Adelberger} K.~L.,  {Dickinson} M.~E.,
  {Pettini} M.,    {Kellogg} M.,  1998, \apj, 503, 543

\bibitem[\protect\citeauthoryear{{Hennawi}, {Strauss}, {Oguri}, {Inada},
  {Richards}, {Pindor}, {Schneider}, {Becker}, {Gregg}, {Hall}, {Johnston},
  {Fan}, {Burles}, {Schlegel}, {Gunn}, {Lupton}, {Bahcall}, {Brunner} \&
  {Brinkmann}}{{Hennawi} et~al.}{2006}]{Hen06}
{Hennawi} J.~F.,  {Strauss} M.~A.,  {Oguri} M.,  {Inada} N.,  {Richards} G.~T.,
   {Pindor} B.,  {Schneider} D.~P.,  {Becker} R.~H.,  {Gregg} M.~D.,  {Hall}
  P.~B.,  {Johnston} D.~E.,  {Fan} X.,  {Burles} S.,  {Schlegel} D.~J.,  {Gunn}
  J.~E.,  {Lupton} R.~H.,  {Bahcall} N.~A.,  {Brunner} R.~J.,    {Brinkmann}
  J.,  2006, \aj, 131, 1

\bibitem[\protect\citeauthoryear{{Hopkins}, {Cox}, {Kere{\v s}} \&
  {Hernquist}}{{Hopkins} et~al.}{2008}]{HopCoxKer08}
{Hopkins} P.~F.,  {Cox} T.~J.,  {Kere{\v s}} D.,    {Hernquist} L.,  2008,
  \apjs, 175, 390

\bibitem[\protect\citeauthoryear{{Hopkins}, {Hernquist}, {Cox} \& {Kere{\v
  s}}}{{Hopkins} et~al.}{2008}]{HopHerCox08}
{Hopkins} P.~F.,  {Hernquist} L.,  {Cox} T.~J.,    {Kere{\v s}} D.,  2008,
  \apjs, 175, 356

\bibitem[\protect\citeauthoryear{{Kashikawa} et~al.,}{{Kashikawa}
  et~al.}{2006}]{Kas06}
{Kashikawa} N.,  et~al., 2006, \apj, 637, 631

\bibitem[\protect\citeauthoryear{{Kitzbichler} \& {White}}{{Kitzbichler} \&
  {White}}{2008}]{KitWhi08}
{Kitzbichler} M.~G.,  {White} S.~D.~M.,  2008, \mnras, 391, 1489

\bibitem[\protect\citeauthoryear{{Klypin}, {Gottl{\"o}ber}, {Kravtsov} \&
  {Khokhlov}}{{Klypin} et~al.}{1999}]{KlyGotKra99}
{Klypin} A.,  {Gottl{\"o}ber} S.,  {Kravtsov} A.~V.,    {Khokhlov} A.~M.,
  1999, \apj, 516, 530

\bibitem[\protect\citeauthoryear{{Komatsu}, {Dunkley}, {Nolta}, {Bennett},
  {Gold}, {Hinshaw}, {Jarosik}, {Larson}, {Limon}, {Page}, {Spergel},
  {Halpern}, {Hill}, {Kogut}, {Meyer}, {Tucker}, {Weiland}, {Wollack} \&
  {Wright}}{{Komatsu} et~al.}{2009}]{WMAP5}
{Komatsu} E.,  {Dunkley} J.,  {Nolta} M.~R.,  {Bennett} C.~L.,  {Gold} B.,
  {Hinshaw} G.,  {Jarosik} N.,  {Larson} D.,  {Limon} M.,  {Page} L.,
  {Spergel} D.~N.,  {Halpern} M.,  {Hill} R.~S.,  {Kogut} A.,  {Meyer} S.~S.,
  {Tucker} G.~S.,  {Weiland} J.~L.,  {Wollack} E.,    {Wright} E.~L.,  2009,
  \apjs, 180, 330

\bibitem[\protect\citeauthoryear{{Kravtsov}, {Berlind}, {Wechsler}, {Klypin},
  {Gottl{\"o}ber}, {Allgood} \& {Primack}}{{Kravtsov}
  et~al.}{2004}]{KraBerWec04}
{Kravtsov} A.~V.,  {Berlind} A.~A.,  {Wechsler} R.~H.,  {Klypin} A.~A.,
  {Gottl{\"o}ber} S.,  {Allgood} B.,    {Primack} J.~R.,  2004, \apj, 609, 35

\bibitem[\protect\citeauthoryear{{Lee}, {Giavalisco}, {Gnedin}, {Somerville},
  {Ferguson}, {Dickinson} \& {Ouchi}}{{Lee} et~al.}{2006}]{Lee06}
{Lee} K.-S.,  {Giavalisco} M.,  {Gnedin} O.~Y.,  {Somerville} R.~S.,
  {Ferguson} H.~C.,  {Dickinson} M.,    {Ouchi} M.,  2006, \apj, 642, 63

\bibitem[\protect\citeauthoryear{{Makino} \& {Hut}}{{Makino} \&
  {Hut}}{1997}]{MakHut97}
{Makino} J.,  {Hut} P.,  1997, \apj, 481, 83

\bibitem[\protect\citeauthoryear{{Masjedi}, {Hogg}, {Cool}, {Eisenstein},
  {Blanton}, {Zehavi}, {Berlind}, {Bell}, {Schneider}, {Warren} \&
  {Brinkmann}}{{Masjedi} et~al.}{2006}]{Mas06}
{Masjedi} M.,  {Hogg} D.~W.,  {Cool} R.~J.,  {Eisenstein} D.~J.,  {Blanton}
  M.~R.,  {Zehavi} I.,  {Berlind} A.~A.,  {Bell} E.~F.,  {Schneider} D.~P.,
  {Warren} M.~S.,    {Brinkmann} J.,  2006, \apj, 644, 54

\bibitem[\protect\citeauthoryear{{Maulbetsch}, {Avila-Reese}, {Col{\'{\i}}n},
  {Gottl{\"o}ber}, {Khalatyan} \& {Steinmetz}}{{Maulbetsch}
  et~al.}{2007}]{Mau07}
{Maulbetsch} C.,  {Avila-Reese} V.,  {Col{\'{\i}}n} P.,  {Gottl{\"o}ber} S.,
  {Khalatyan} A.,    {Steinmetz} M.,  2007, \apj, 654, 53

\bibitem[\protect\citeauthoryear{{Moore}, {Ghigna}, {Governato}, {Lake},
  {Quinn}, {Stadel} \& {Tozzi}}{{Moore} et~al.}{1999}]{Moo99}
{Moore} B.,  {Ghigna} S.,  {Governato} F.,  {Lake} G.,  {Quinn} T.,  {Stadel}
  J.,    {Tozzi} P.,  1999, \apjl, 524, L19

\bibitem[\protect\citeauthoryear{{Myers}, {Richards}, {Brunner}, {Schneider},
  {Strand}, {Hall}, {Blomquist} \& {York}}{{Myers} et~al.}{2008}]{Mye08}
{Myers} A.~D.,  {Richards} G.~T.,  {Brunner} R.~J.,  {Schneider} D.~P.,
  {Strand} N.~E.,  {Hall} P.~B.,  {Blomquist} J.~A.,    {York} D.~G.,  2008,
  \apj, 678, 635

\bibitem[\protect\citeauthoryear{{Nagai} \& {Kravtsov}}{{Nagai} \&
  {Kravtsov}}{2005}]{NagKra05}
{Nagai} D.,  {Kravtsov} A.~V.,  2005, \apj, 618, 557

\bibitem[\protect\citeauthoryear{{Navarro}, {Frenk} \& {White}}{{Navarro}
  et~al.}{1996}]{NFW96}
{Navarro} J.~F.,  {Frenk} C.~S.,    {White} S.~D.~M.,  1996, \apj, 462, 563

\bibitem[\protect\citeauthoryear{{Noguchi}}{{Noguchi}}{1991}]{Nog91}
{Noguchi} M.,  1991, \mnras, 251, 360

\bibitem[\protect\citeauthoryear{{Ouchi}, {Hamana}, {Shimasaku}, {Yamada},
  {Akiyama}, {Kashikawa}, {Yoshida}, {Aoki}, {Iye}, {Saito}, {Sasaki},
  {Simpson} \& {Yoshida}}{{Ouchi} et~al.}{2005}]{Ouc05}
{Ouchi} M.,  {Hamana} T.,  {Shimasaku} K.,  {Yamada} T.,  {Akiyama} M.,
  {Kashikawa} N.,  {Yoshida} M.,  {Aoki} K.,  {Iye} M.,  {Saito} T.,  {Sasaki}
  T.,  {Simpson} C.,    {Yoshida} M.,  2005, \apjl, 635, L117

\bibitem[\protect\citeauthoryear{{Padmanabhan}, {White}, {Norberg} \&
  {Porciani}}{{Padmanabhan} et~al.}{2008}]{Pad08}
{Padmanabhan} N.,  {White} M.,  {Norberg} P.,    {Porciani} C.,  2008, arXiv:
  0802.2105

\bibitem[\protect\citeauthoryear{{Peacock} \& {Smith}}{{Peacock} \&
  {Smith}}{2000}]{PeaSmi00}
{Peacock} J.~A.,  {Smith} R.~E.,  2000, \mnras, 318, 1144

\bibitem[\protect\citeauthoryear{{Percival}, {Scott}, {Peacock} \&
  {Dunlop}}{{Percival} et~al.}{2003}]{Per03}
{Percival} W.~J.,  {Scott} D.,  {Peacock} J.~A.,    {Dunlop} J.~S.,  2003,
  \mnras, 338, L31

\bibitem[\protect\citeauthoryear{{Reichardt}, {Ade}, {Bock}, {Bond}, {Brevik},
  {Contaldi}, {Daub}, {Dempsey}, {Goldstein}, {Holzapfel}, {Kuo}, {Lange},
  {Lueker}, {Newcomb}, {Peterson}, {Ruhl}, {Runyan} \&
  {Staniszewski}}{{Reichardt} et~al.}{2008}]{ACBAR}
{Reichardt} C.~L.,  {Ade} P.~A.~R.,  {Bock} J.~J.,  {Bond} J.~R.,  {Brevik}
  J.~A.,  {Contaldi} C.~R.,  {Daub} M.~D.,  {Dempsey} J.~T.,  {Goldstein}
  J.~H.,  {Holzapfel} W.~L.,  {Kuo} C.~L.,  {Lange} A.~E.,  {Lueker} M.,
  {Newcomb} M.,  {Peterson} J.~B.,  {Ruhl} J.,  {Runyan} M.~C.,
  {Staniszewski} Z.,  2008, arXiv: 0801.1491

\bibitem[\protect\citeauthoryear{{Sanders} \& {Mirabel}}{{Sanders} \&
  {Mirabel}}{1996}]{SanMir96}
{Sanders} D.~B.,  {Mirabel} I.~F.,  1996, \araa, 34, 749

\bibitem[\protect\citeauthoryear{{Saro}, {De Lucia}, {Dolag} \&
  {Borgani}}{{Saro} et~al.}{2008}]{Sar08}
{Saro} A.,  {De Lucia} G.,  {Dolag} K.,    {Borgani} S.,  2008, \mnras, 391,
  565

\bibitem[\protect\citeauthoryear{{Scannapieco} \& {Thacker}}{{Scannapieco} \&
  {Thacker}}{2003}]{ScaTha03}
{Scannapieco} E.,  {Thacker} R.~J.,  2003, \apjl, 590, L69

\bibitem[\protect\citeauthoryear{{Scott}, {Dunlop} \& {Serjeant}}{{Scott}
  et~al.}{2006}]{ScoDunSer06}
{Scott} S.~E.,  {Dunlop} J.~S.,    {Serjeant} S.,  2006, \mnras, 370, 1057

\bibitem[\protect\citeauthoryear{{Seljak}}{{Seljak}}{2000}]{Sel00}
{Seljak} U.,  2000, \mnras, 318, 203

\bibitem[\protect\citeauthoryear{{Shen}, {Strauss}, {Oguri}, {Hennawi}, {Fan},
  {Richards}, {Hall}, {Gunn}, {Schneider}, {Szalay}, {Thakar}, {Vanden Berk},
  {Anderson}, {Bahcall}, {Connolly} \& {Knapp}}{{Shen} et~al.}{2007}]{She07}
{Shen} Y.,  {Strauss} M.~A.,  {Oguri} M.,  {Hennawi} J.~F.,  {Fan} X.,
  {Richards} G.~T.,  {Hall} P.~B.,  {Gunn} J.~E.,  {Schneider} D.~P.,  {Szalay}
  A.~S.,  {Thakar} A.~R.,  {Vanden Berk} D.~E.,  {Anderson} S.~F.,  {Bahcall}
  N.~A.,  {Connolly} A.~J.,    {Knapp} G.~R.,  2007, \aj, 133, 2222

\bibitem[\protect\citeauthoryear{{Sheth} \& {Tormen}}{{Sheth} \&
  {Tormen}}{2004}]{SheTor04}
{Sheth} R.~K.,  {Tormen} G.,  2004, \mnras, 350, 1385

\bibitem[\protect\citeauthoryear{{Smoot} et~al.,}{{Smoot}  et~al.}{1992}]{COBE}
{Smoot} G.~F.,  et~al., 1992, \apjl, 396, L1

\bibitem[\protect\citeauthoryear{{Springel}, {Di Matteo} \&
  {Hernquist}}{{Springel} et~al.}{2005}]{SprDiMHer05b}
{Springel} V.,  {Di Matteo} T.,    {Hernquist} L.,  2005, \mnras, 361, 776

\bibitem[\protect\citeauthoryear{{Springel}, {White}, {Jenkins}, {Frenk},
  {Yoshida}, {Gao}, {Navarro}, {Thacker}, {Croton}, {Helly}, {Peacock}, {Cole},
  {Thomas}, {Couchman}, {Evrard}, {Colberg} \& {Pearce}}{{Springel}
  et~al.}{2005}]{Spr05}
{Springel} V.,  {White} S.~D.~M.,  {Jenkins} A.,  {Frenk} C.~S.,  {Yoshida} N.,
   {Gao} L.,  {Navarro} J.,  {Thacker} R.,  {Croton} D.,  {Helly} J.,
  {Peacock} J.~A.,  {Cole} S.,  {Thomas} P.,  {Couchman} H.,  {Evrard} A.,
  {Colberg} J.,    {Pearce} F.,  2005, \nat, 435, 629

\bibitem[\protect\citeauthoryear{{Springel}, {White}, {Tormen} \&
  {Kauffmann}}{{Springel} et~al.}{2001}]{Spr01}
{Springel} V.,  {White} S.~D.~M.,  {Tormen} G.,    {Kauffmann} G.,  2001,
  \mnras, 328, 726

\bibitem[\protect\citeauthoryear{{Tegmark} et~al.,}{{Tegmark}
  et~al.}{2006}]{Teg06}
{Tegmark} M.,  et~al., 2006, \prd, 74, 123507

\bibitem[\protect\citeauthoryear{{Thacker}, {Scannapieco} \&
  {Couchman}}{{Thacker} et~al.}{2006}]{ThaScaCou06}
{Thacker} R.~J.,  {Scannapieco} E.,    {Couchman} H.~M.~P.,  2006, \apj, 653,
  86

\bibitem[\protect\citeauthoryear{{Tinker}, {Conroy}, {Norberg}, {Patiri},
  {Weinberg} \& {Warren}}{{Tinker} et~al.}{2008}]{Tin08}
{Tinker} J.~L.,  {Conroy} C.,  {Norberg} P.,  {Patiri} S.~G.,  {Weinberg}
  D.~H.,    {Warren} M.~S.,  2008, \apj, 686, 53

\bibitem[\protect\citeauthoryear{{Tormen}}{{Tormen}}{1997}]{Tor97}
{Tormen} G.,  1997, \mnras, 290, 411

\bibitem[\protect\citeauthoryear{{Tormen}, {Diaferio} \& {Syer}}{{Tormen}
  et~al.}{1998}]{TorDiaSye98}
{Tormen} G.,  {Diaferio} A.,    {Syer} D.,  1998, \mnras, 299, 728

\bibitem[\protect\citeauthoryear{{Vale} \& {Ostriker}}{{Vale} \&
  {Ostriker}}{2006}]{ValOst06}
{Vale} A.,  {Ostriker} J.~P.,  2006, \mnras, 371, 1173

\bibitem[\protect\citeauthoryear{{Wang}, {Li}, {Kauffmann} \& {De
  Lucia}}{{Wang} et~al.}{2006}]{WanKauDeL06}
{Wang} L.,  {Li} C.,  {Kauffmann} G.,    {De Lucia} G.,  2006, \mnras, 371, 537

\bibitem[\protect\citeauthoryear{{Wechsler}, {Zentner}, {Bullock}, {Kravtsov}
  \& {Allgood}}{{Wechsler} et~al.}{2006}]{Wec06}
{Wechsler} R.~H.,  {Zentner} A.~R.,  {Bullock} J.~S.,  {Kravtsov} A.~V.,
  {Allgood} B.,  2006, \apj, 652, 71

\bibitem[\protect\citeauthoryear{{Wetzel}, {Cohn} \& {White}}{{Wetzel}
  et~al.}{2009}]{WetCohWhi09a}
{Wetzel} A.~R.,  {Cohn} J.~D.,    {White} M.,  2009, \mnras, p. in press

\bibitem[\protect\citeauthoryear{{Wetzel}, {Cohn}, {White}, {Holz} \&
  {Warren}}{{Wetzel} et~al.}{2007}]{Wet07}
{Wetzel} A.~R.,  {Cohn} J.~D.,  {White} M.,  {Holz} D.~E.,    {Warren} M.~S.,
  2007, \apj, 656, 139

\bibitem[\protect\citeauthoryear{{Wetzel}, {Schulz}, {Holz} \&
  {Warren}}{{Wetzel} et~al.}{2008}]{WetSchHol08}
{Wetzel} A.~R.,  {Schulz} A.~E.,  {Holz} D.~E.,    {Warren} M.~S.,  2008, \apj,
  683, 1

\bibitem[\protect\citeauthoryear{{White}}{{White}}{2002}]{TreePM}
{White} M.,  2002, \apjs, 143, 241

\bibitem[\protect\citeauthoryear{{Yamauchi}, {Yagi} \& {Goto}}{{Yamauchi}
  et~al.}{2008}]{YamYagGot08}
{Yamauchi} C.,  {Yagi} M.,    {Goto} T.,  2008, \mnras, 390, 383

\bibitem[\protect\citeauthoryear{{Yan}, {Newman}, {Faber}, {Coil}, {Cooper},
  {Davis}, {Weiner}, {Gerke} \& {Koo}}{{Yan} et~al.}{2008}]{Yan08}
{Yan} R.,  {Newman} J.~A.,  {Faber} S.~M.,  {Coil} A.~L.,  {Cooper} M.~C.,
  {Davis} M.,  {Weiner} B.~J.,  {Gerke} B.~F.,    {Koo} D.~C.,  2008, arXiv:
  0805.0004

\bibitem[\protect\citeauthoryear{{Yang}, {Mo} \& {van den Bosch}}{{Yang}
  et~al.}{2009}]{YanMovdB08}
{Yang} X.,  {Mo} H.~J.,    {van den Bosch} F.~C.,  2009, \apj, 693, 830

\bibitem[\protect\citeauthoryear{{Yoshida}, {Shimasaku}, {Ouchi}, {Sekiguchi},
  {Furusawa} \& {Okamura}}{{Yoshida} et~al.}{2008}]{Yos08}
{Yoshida} M.,  {Shimasaku} K.,  {Ouchi} M.,  {Sekiguchi} K.,  {Furusawa} H.,
  {Okamura} S.,  2008, \apj, 679, 269

\bibitem[\protect\citeauthoryear{{Zentner}, {Berlind}, {Bullock}, {Kravtsov} \&
  {Wechsler}}{{Zentner} et~al.}{2005}]{ZenBerBul05}
{Zentner} A.~R.,  {Berlind} A.~A.,  {Bullock} J.~S.,  {Kravtsov} A.~V.,
  {Wechsler} R.~H.,  2005, \apj, 624, 505

\end{thebibliography}

\label{lastpage}

\end{document}